\documentclass[12pt]{article}
\usepackage{physics}
\usepackage{epsf}
\usepackage{aas_macros}
\usepackage{amssymb}
\usepackage{comment}
\usepackage{amsmath}
\usepackage{braket}
\usepackage{mathrsfs}
\usepackage{mathtools}
\usepackage{array}
\usepackage{fancyhdr}
\usepackage[dvipdfmx]{graphicx}
\usepackage{color}
\usepackage{cite}
\usepackage{bm}
\usepackage{url}
\usepackage{fancyhdr}
\usepackage[colorlinks,citecolor=blue]{hyperref}
\usepackage[hang,small,bf]{caption}
\usepackage[subrefformat=parens]{subcaption}
\usepackage{scalerel}
\usepackage[usestackEOL]{stackengine}

\renewcommand{\thefootnote}{\#\arabic{footnote}}

\renewcommand{\thefootnote}{\fnsymbol{footnote}}
\def\thefootnote{\fnsymbol{footnote}}

\makeatletter

\@addtoreset{equation}{section}
\makeatother

\def\be{\begin{equation}}
\def\ee{\end{equation}}
\def\ben{\begin{eqnarray}}
\def\een{\end{eqnarray}}

\begin{document}


\begin{center}

\vskip .75in

{\Large \bf Novel Quantity for Probing Matter Perturbations Below the Fresnel Scale in Gravitational Lensing of Gravitational Waves}

\vskip .75in

{\large
So Tanaka$\,^1$,  Teruaki Suyama$\,^1$
}

\vskip 0.25in

{\em
$^{1}$Department of Physics, Institute of Science Tokyo, 2-12-1 Ookayama, Meguro-ku,
Tokyo 152-8551, Japan
}

\end{center}
\vskip .5in

\begin{abstract}
Gravitational lensing of gravitational waves provides a powerful probe of the mass density distribution in the universe.
Wave optics effects, such as diffraction, make the lensing effect sensitive to the structure around the Fresnel scale, which depends on the gravitational wave frequency and is typically sub-Galactic for realistic observations.
Contrary to this common lore, we show that wave optics can, in principle, probe matter perturbations even below the Fresnel scale.
This is achieved by introducing a new quantity derived from the amplification factor, which characterizes the lensing effect, and analyzing its correlation function.
Our results demonstrate that this quantity defines an effective Fresnel scale—a characteristic scale that can be arbitrarily small, even when observational frequencies are bounded.
In practice, the effective Fresnel scale is constrained by the observation time $T$ and is suppressed by a factor of $1/\sqrt{fT}$ relative to the standard Fresnel scale at frequency $f$. 
Nevertheless, it remains significantly smaller than the conventional Fresnel scale for $fT \gg 1$; for instance, in one-year observations of mHz GWs, the effective Fresnel scale can be as small as 1 pc. 
This approach opens new avenues for probing the fine-scale structure of the universe and the nature of dark matter.
\end{abstract}

\renewcommand{\thepage}{\arabic{page}}
\setcounter{page}{1}
\renewcommand{\thefootnote}{\#\arabic{footnote}}
\setcounter{footnote}{0}

\section{Introduction}\label{Sec: introduction}

Gravitational lensing is a powerful tool for probing the mass density distribution in the universe\cite{Oguri:2020ldf,Savastano:2023spl,Takahashi:2005ug,Tambalo:2022wlm}.
In particular, the small-scale structure below the Mpc scale remains largely unknown yet, and revealing its properties is crucial for understanding the nature of dark matter\cite{Moore:1999nt,Lin:2019uvt}.
While only the gravitational lensing of light has been observed so far, gravitational lensing of gravitational waves (GWs) has garnered significant attention in recent years\cite{Choi:2021bkx,Tambalo:2022plm,Dai:2018enj,Tanaka:2023mvy}.
Due to the much longer typical wavelength of GWs compared to light, wave optics effects play a crucial role in their propagation for some lens systems\cite{Nakamura:1997sw, Nakamura:1999uwi, Takahashi:2003ix}. 
Diffraction is one such effect, where GWs pass through extended regions, enabling them to probe the mass density distribution.
In addition, the diffraction effect is frequency-dependent, allowing us to explore the structure at different scales\cite{Takahashi:2005ug, Kim:2025njb}.

In this paper, we examine the lensing of low-frequency GWs by randomly distributed matter fluctuations.
For such waves, the lensing effect is weak, meaning the lensed waveform deviates only slightly from the unlensed one.
In the weak lensing regime, it is a good approximation to compute the lensing effect to first order in the gravitational potential of the lensing objects (the Born approximation), which we adopt throughout this study\cite{Mizuno:2022xxp}.
In \cite{Takahashi:2005ug}, it was shown that the statistical fluctuation of the amplification factor is sensitive to the structure with a scale larger than the Fresnel scale $r_F$, which depends on frequency.
This suggests that by analyzing how the amplification factor varies with frequency, we can extract information about the matter power spectrum around the Fresnel scale.
Since the typical value of the Fresnel scale is $r_F\sim 100\,\text{pc}\,(f/\text{mHz})^{-1/2}(D_s/10\,\text{Gpc})^{1/2}$, where $f$ is the frequency of GWs and $D_s$ is the distance from the observer to the wave source, the statistical degrees of fluctuation of the amplification factor is a promising observable for probing the unknown small-scale structure.

In this work, we introduce a new quantity derived from the amplification factor in Sec. \ref{Sec: new quantity formalization}.
We then calculate its correlation function between different frequencies, showing that it is more sensitive to the small-scale structure than the amplification factor itself. 
Specifically, we show that it is possible to probe the structure at an arbitrarily small scale, even when there is an upper limit to the frequency.
We also discuss the observability of our new quantity in Sec. \ref{Sec: new quantity observability}.
Throughout this study, we adopt the unit $c=1$.


\section{Amplification Factor in the Born Approximation}\label{Sec: review}

In this section, we review the formalism to compute the amplification factor in the Born approximation\cite{Takahashi:2005sxa}. 
The background metric $\tilde{g}^B_{\mu\nu}$ is the flat FLRW metric perturbed by the Newtonian potential $\Phi$:
\begin{eqnarray}
    ds^2
    &=&\tilde{g}^B_{\mu\nu}dx^{\mu}dx^{\nu}\nonumber\\
    &=&a^2g^B_{\mu\nu}dx^{\mu}dx^{\nu}\nonumber\\
    &=&a^2(\eta)\qty[-(1+2\Phi(\bm{r};\eta))d\eta^2+(1-2\Phi(\bm{r};\eta))d\bm{r}^2]\, ,
\end{eqnarray}
where $\eta$ is conformal time$,\bm{r}$ are comoving coordinates, 
$a(\eta)$ is the scale factor, and $g^B_{\mu\nu}$ is the perturbed Minkowski metric.
Since we are considering the situation where the period of GWs is much smaller than the timescale of the perturbations, we ignore the time dependence of $\Phi$ in the following calculations and then reconsider it later.

Let us denote GWs as $\tilde{h}_{\mu\nu}=a^2h_{\mu\nu}$.
GWs have amplitude and polarization, i.e. $h_{\mu\nu}=\phi e_{\mu\nu}$, but we treat it as a scalar wave $\phi$ since the change in polarization during the propagation is small\cite{peters1974index}. 
In addition, the wave equation is conformally invariant when the wavelength is much smaller than the Hubble radius\cite{Takahashi:2005ug}. 
Thus, the propagation of the GW is described by the wave equation for the scalar wave $\phi$ on the perturbed Minkowski background $g^B_{\mu\nu}$:
\begin{equation}
    \partial_{\mu}\qty[\sqrt{-g^B}g_B^{\mu\nu}\partial_{\nu}\phi]=0
\end{equation}
or 
\begin{equation}
    (\nabla^2+\omega^2)\tilde{\phi}(\omega,\bm{r})=4\omega^2\Phi(\bm{r})\tilde{\phi}(\omega,\bm{r})\, ,\label{Eq: wave eq}
\end{equation}
where $\omega$ is the comoving frequency (frequency at the observer) and $\tilde{\phi}(\omega,\bm{r})$ is the Fourier transform of $\phi(\eta,\bm{r})$, and the higher order terms of $\Phi$ are neglected.

\begin{figure}[t]
    \centering
    \includegraphics[scale=0.3]{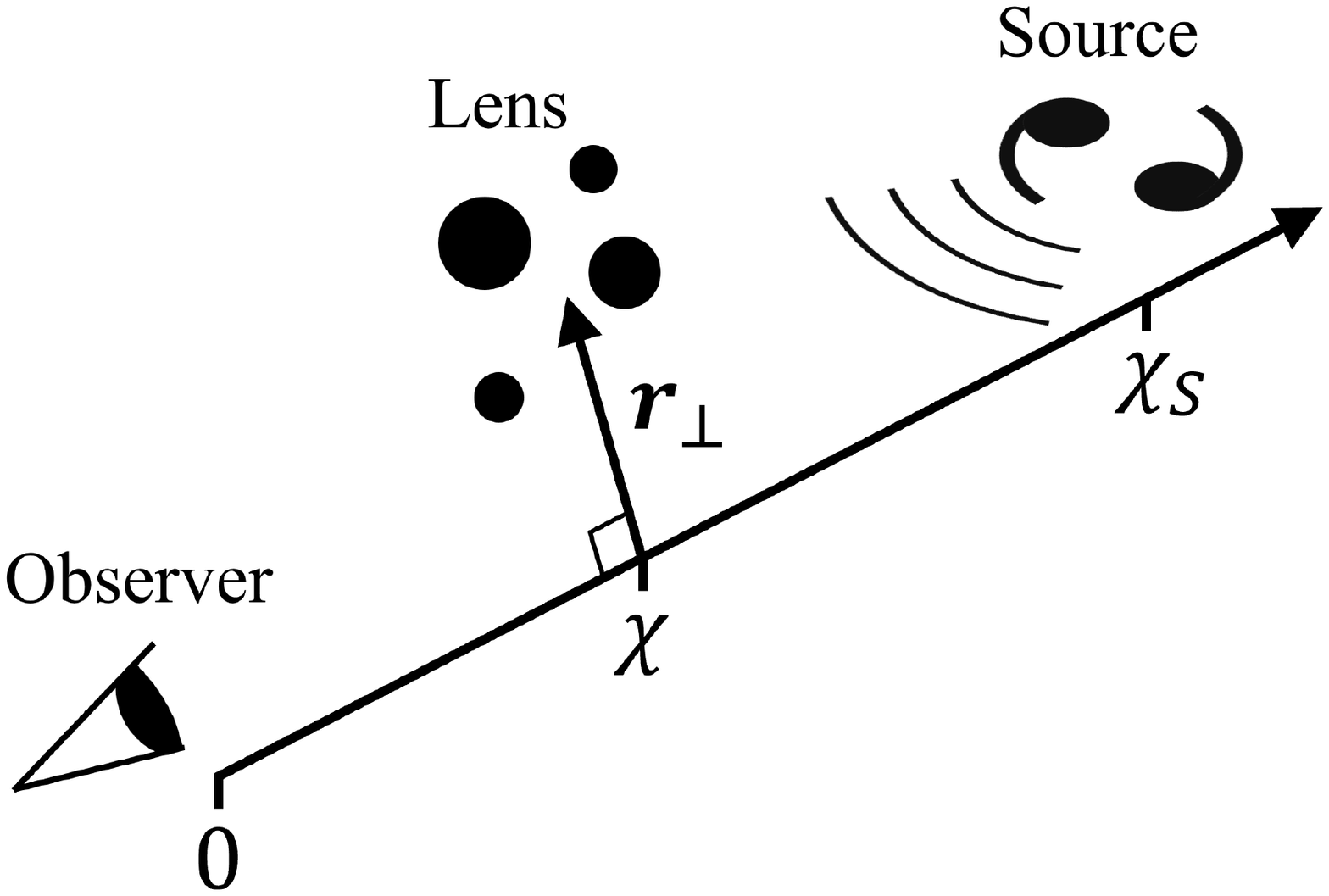}
    \caption{Schematic picture of gravitational lensing.
    $\chi$-axis is directed from the observer to the wave source, and the source is at $\chi=\chi_s.$
    $\bm{r}_\perp{}$ is the position vector perpendicular to the $\chi$-axis.}
    \label{Fig: schematic picture}
\end{figure}

When the right-hand side of Eq. (\ref{Eq: wave eq}) only acts as a small perturbation, which is the case if $\omega$ is small, the Born approximation is valid\cite{Mizuno:2022xxp}.
To solve the wave equation (\ref{Eq: wave eq}) at the leading order of $\Phi$, we decompose the solution as $\tilde{\phi}=\tilde{\phi}_0+\tilde{\phi_1}$, where $\tilde{\phi}_0$ is the unlensed spherical wave emitted from the source at $\bm{r}=\bm{r}_s$:
\begin{equation}
    \tilde{\phi}_0(\omega,\bm{r})=\frac{e^{i\omega|\bm{r}-\bm{r}_s|}}{|\bm{r}-\bm{r}_s|}\,,
\end{equation}
and $\tilde{\phi}_1$ is the solution of 
\begin{equation}
     (\nabla^2+\omega^2)\tilde{\phi}_1(\omega,\bm{r})=4\omega^2\Phi(\bm{r})\tilde{\phi}_0(\omega,\bm{r})\, .\label{Eq: wave eq perturbation}
\end{equation}
Then the amplification factor is defined by
\begin{eqnarray}
    F(\omega)
    =\frac{\tilde{\phi}(\omega,\bm{r}_o)}{\tilde{\phi}_0(\omega,\bm{r}_o)}
    =1+\frac{\tilde{\phi}_1(\omega,\bm{r}_o)}{\tilde{\phi}_0(\omega,\bm{r}_o)}\, ,
\end{eqnarray}
where $\bm{r}_o$ is the position of the observer and we set $\bm{r}_o=\bm{0}$ in the following.
From Eq. (\ref{Eq: wave eq perturbation}), we obtain \cite{Takahashi:2005sxa}
\begin{equation}
    F(\omega)-1=-\frac{\omega^2}{\pi}\int d^3r\,\Phi(\bm{r})\frac{|\bm{r}_s|}{|\bm{r}_s-\bm{r}||\bm{r}|}e^{i\omega(|\bm{r}_s-\bm{r}|+|\bm{r}|-|\bm{r}_s|)}\,.\label{Eq: ampfac position intermediate}
\end{equation}
Then we choose the $\chi$-axis along $\bm{r}_s$ and write $\bm{r}=(\bm{r}_\perp,\chi)$ and $\bm{r}_s=(\bm{0},\chi_s)$, where $\chi_s$ is the comoving distance from the observer to the source (see Fig. \ref{Fig: schematic picture}). 
Assuming that only the vicinity of the line of sight contributes ($r_{\perp}\ll \chi,\,\chi_s-\chi$), Eq. (\ref{Eq: ampfac position intermediate}) simplifies to
\begin{eqnarray}
    F(\omega)-1
    &=&-\frac{\omega^2}{\pi}\int_0^{\chi_s} d\chi\int d^2r_{\perp}\,\Phi(\bm{r})\frac{1}{d_{\rm eff}(\chi)}\exp\qty[i\omega\frac{r_{\perp}^2}{2d_{\rm eff}(\chi)}]\,,\label{Eq: ampfac position}
\end{eqnarray}
where $d_{\rm eff}$ is the effective coordinate distance defined by
\begin{equation}
    d_{\rm eff}(\chi)=\frac{\chi(\chi_s-\chi)}{\chi_s}\,.
\end{equation}
Here we reconsider the time dependence of $\Phi$.
In Eq. (\ref{Eq: ampfac position}), $\Phi(\bm{r})$ should be replaced by $\Phi(\bm{r};\eta(\chi))$, where $\eta(\chi)$ is the time when the GWs pass through the plane at a distance $\chi$. 
By setting $\eta=0$ at the time when the GWs departed from the source and redefining $\chi$ as the distance from the source rather than from the observer ($\chi_s-\chi\rightarrow \chi$), we obtain $\eta(\chi)=\chi$. 
Then Eq. (\ref{Eq: ampfac position}) becomes
\begin{equation}
    F(\omega)-1=-2i\omega\int_0^{\chi_s} d\chi\int \frac{d^2k_{\perp}}{(2\pi)^2}\,\tilde{\Phi}(\bm{k}_{\perp},\chi;\chi)\exp\qty[-\frac{i}{2}r_F^2(\omega,\chi)k_\perp^2]\,,\label{Eq: ampfac momentum intermediate}
\end{equation}
where we have defined the Fresnel scale $r_F$ by
\begin{equation}
    r_F(\omega,\chi)=\sqrt{\frac{d_{\rm eff}(\chi)}{\omega}}
\end{equation}
and $\tilde{\Phi}(\bm{k}_\perp,\chi;\eta)$ is the Fourier transform of $\Phi(\bm{r}_\perp,\chi;\eta)$ with respect to $\bm{r}_\perp$.

To obtain a more specific expression for $F(\omega)$, we use the Poisson equation satisfied by $\Phi$\cite{Peebles:1980yev}:
\begin{equation}
    \nabla^2\Phi(\bm{r};\eta)=4\pi G a^2(\eta)\delta\rho(\bm{r};\eta)\,,
\end{equation}
where $\delta\rho$ is the mass density perturbations.
Taking the Fourier transform, we obtain
\begin{equation}
    \tilde{\Phi}(\bm{k};\eta)=-4\pi Ga^2(\eta)\frac{\delta\tilde{\rho}(\bm{k};\eta)}{k^2}\,,\label{Eq: sol poisson momentum}
\end{equation}
where $\tilde{\Phi}(\bm{k};\eta)$ and $\delta\tilde{\rho}(\bm{k};\eta)$ are the Fourier transforms of $\Phi(\bm{r};\eta)$ and $\delta\rho(\bm{r};\eta)$, respectively.
As can be seen from Eq. (\ref{Eq: ampfac position}), the typical coordinate scales contributing to the integral are $r_\perp\sim r_F\sim (\chi_s/\omega)^{1/2}$ and $\chi\sim \chi_s$.
Consequently, in momentum space, the characteristic scales are $k_\perp\sim(\omega/\chi_s)^{1/2}$ and $k_\parallel\sim 1/\chi_s$.
Assuming $\omega\gg1/\chi_s$, it follows $k_\perp\gg k_\parallel$, and the approximation $k \approx k_\perp$ (Limber approximation) in Eq. (\ref{Eq: sol poisson momentum}) gives
\begin{equation}
    \tilde{\Phi}(\bm{k}_\perp,\chi;\eta)=-4\pi Ga^2(\eta)\frac{\delta\tilde{\rho}(\bm{k}_\perp,\chi;\eta)}{k_\perp^2}\,.\label{Eq: sol poisson momentum z}
\end{equation}
By substituting this into Eq. (\ref{Eq: ampfac momentum intermediate}) and defining the critical surface mass density as
\begin{equation}
    \Sigma_{c}(\chi) = \frac{1}{4\pi G d_{\rm eff}(\chi)a^2(\chi)}\,,
\end{equation}
we obtain
\begin{equation}
    F(\omega)-1=\int_0^{\chi_s} d\chi\int \frac{d^2k_{\perp}}{(2\pi)^2}\,\frac{\delta\tilde{\rho}(\bm{k}_\perp,\chi;\chi)}{\Sigma_c(\chi)}\frac{e^{-ir_F^2k_\perp^2/2}}{-ir_F^2k_\perp^2/2}\,.
\end{equation}
Large frequency limit ($\omega \rightarrow\infty$) of the right-hand side of the above equation corresponds to the Shapiro time delay: $t_s=-2\int_0^{\chi_s}d\chi\,\Phi(\bm{0},\chi)$.
Since it is not a directly observable quantity\cite{Takahashi:2005ug}, we redefine $F(\omega)$ by removing the Shapiro time delay:
\begin{equation}
    F(\omega)-1=\int_0^{\chi_s} d\chi\int \frac{d^2k_{\perp}}{(2\pi)^2}\,\frac{\delta\tilde{\rho}(\bm{k}_\perp,\chi;\chi)}{\Sigma_c(\chi)}W\qty(-\frac{i}{2}r_F^2(\omega,\chi)k_\perp^2)\,,\label{Eq: ampfac momentum}
\end{equation}
where the Window function $W(x)$ is given by
\begin{equation}
    W(x)=\frac{e^{x}-1}{x}\,.
\end{equation}
Due to the Window function, modes with $k_\perp>1/r_F(\omega,\chi)$ are suppressed. 
Physically, this can be understood as follows: when the separation between two optical paths from the source to the observer is smaller than the Fresnel scale, their path difference is shorter than the wavelength of GWs.
Consequently, these paths contribute equally to the lensing effect, meaning that matter perturbations smoothed over the Fresnel scale contribute to the amplification factor.
As a result, $F(\omega)$ is sensitive only to the structure with $k_\perp<1/r_F(\omega,\chi)$ and by measuring how $F(\omega)$ varies with $\omega$, we can probe the matter perturbations around the Fresnel scale. 
The detectability of the lens signal is discussed in \cite{Oguri:2020ldf, Mizuno:2022xxp}.
From the above argument, it seems that it is not possible to probe the perturbations below the Fresnel scale from the measurement of the lensed waveform.
However, as we will discuss in Sec. \ref{Sec: new quantity formalization}, a new quantity we propose allows us to probe the small-scale structure below the Fresnel scale using only low-frequency information.


\section{A Novel Quantity for Probing Matter Perturbations Below
the Fresnel Scale}


\subsection{Formulation}\label{Sec: new quantity formalization}

In this subsection, we introduce a new quantity derived from the amplification factor and demonstrate that it is sensitive to structures smaller than the Fresnel scale.
To this end we define $I(\omega)$ by
\footnote{
A related (though not identical) quantity was introduced in \cite{Choi:2021bkx}, and the connection between the two is discussed in Appendix \ref{appendix-A}.
}
\begin{equation}
    I(\omega)=-\omega^2\frac{d}{d\omega}\frac{F(\omega)-1}{\omega}\label{Eq: new quantity def}
\end{equation}
and from Eq. (\ref{Eq: ampfac momentum}), we obtain
\begin{equation}
    I(\omega)=\int_0^{\chi_s} d\chi\int \frac{d^2k_{\perp}}{(2\pi)^2}\,\frac{\delta\tilde{\rho}(\bm{k}_\perp,\chi;\chi)}{\Sigma_c(\chi)}e^{-ir_F^2k_\perp^2/2}\,.
\end{equation}
To compute the correlation function of $I(\omega)$, we first examine the correlation function of $\delta\rho$.
Assuming statistical homogeneity and isotropy of the matter perturbations, and using the power spectrum $P_{\delta\rho}$, we have 
\begin{equation}
    \langle\delta\tilde{\rho}^*(\bm{k};\eta)\delta\tilde{\rho}(\bm{k}';\eta)\rangle=(2\pi)^3\delta^3(\bm{k}-\bm{k}')P_{\delta\rho}(k;\eta)\,,
\end{equation}
where $\langle\rangle$ denotes the ensemble average, and then
\begin{equation}
    \langle\delta\tilde{\rho}^*(\bm{k}_\perp,\chi;\eta)\delta\tilde{\rho}(\bm{k}'_\perp,\chi';\eta)\rangle
    =(2\pi)^2\delta^2(\bm{k}_\perp-\bm{k}_\perp')\delta(\chi-\chi')P_{\delta\rho}(k_\perp;\eta)\,,
\end{equation}
where we have approximated $k\approx k_\perp$ based on $k_\perp\gg k_\parallel$, as described above in Eq. (\ref{Eq: sol poisson momentum z}).
Then we find the correlation function between $I(\omega)$ and $I(\omega')$ is given by
\begin{equation}
    \Delta_I(\Omega)=
    \langle I^*(\omega)I(\omega')\rangle=\int_0^{\chi_s} d\chi\int \frac{d^2k_{\perp}}{(2\pi)^2}\,\frac{P_{\delta\rho}(k_\perp;\chi)}{\Sigma_c^2(\chi)}\exp\qty[-\frac{i}{2}r_F^2(\Omega,\chi)k_\perp^2]\,,\label{Eq: correlation}
\end{equation}
where $\Omega$ is a new frequency defined by $\Omega=\omega\omega'/(\omega-\omega')$.
We observe that the new Fresnel scale defined by the frequency $\Omega$ enters this correlation function.
By the same reasoning as discussed at the end of Sec. \ref{Sec: review}, modes with $k_\perp>1/r_F(\Omega, \chi)$ are suppressed, making $\Delta_I(\Omega)$ sensitive only to the structure with $k_\perp<1/r_F(\Omega, \chi)$. 
However, the key difference between $\Delta_I(\Omega)$ and $F(\omega)$ is that $\Omega$ can take an arbitrarily large value by bringing $\omega$ and $\omega'$ closer together, even when $\omega$ and $\omega'$ have an upper bound.
Specifically, if $\omega_{\rm min}\leq|\omega|\leq\omega_{\rm max}$
\footnote{
Negative frequency is defined by $F(-\omega)=F^*(\omega)$, which follows from the requirement that the Fourier transform of $F(\omega)$ must be real.
}
, then $\omega_{\rm min}/2\leq|\Omega|<\infty$.
Consequently, $r_F(\Omega, \chi)$ can become arbitrarily small, and thus $\Delta_I(\Omega)$ enables us to probe the structure at an arbitrarily small scale, even when there is an observational upper bound on the frequency range.

In reality, due to the finite observation time $T$, the difference between two frequencies $\omega$ and $\omega'$ can only be as small as $|\omega-\omega'| \gtrsim 1/T$.
Thus, setting $\omega'=\omega-1/T$ ($\omega \gg 1/T$), the realistic minimum Fresnel scale for neighboring frequencies is given by
\begin{equation}
    r_{F,{\rm min}} (\Omega,\chi) \simeq \frac{1}{\sqrt{\omega T}} r_F(\omega,\chi)\,.
\end{equation}
For $\omega T \gg 1$, the Fresnel scale on the left-hand side is significantly suppressed compared to the original Fresnel scale $r_F (\omega, \chi)$.
To get the flavor, let us consider a one-year observation of ${\cal O}({\rm mHz})$ GWs coming from the cosmological distance
\footnote{
An equal-mass circular binary emitting GWs at frequency $f$ will coalesce in approximately $20\,{\rm yr}\,\qty(M_c/10^3\,M_\odot)^{-5/3}\qty(f/1\,{\rm mHz})^{-8/3}$, where $M_c$ is the chirp mass.
}
. Then, we have
\begin{equation}
    r_{F,{\rm min}} (\Omega,\chi) \simeq 1\,{\rm pc}\, {\qty( \frac{f}{1\,\rm mHz})}^{-1}
    {\qty( \frac{T}{{1\,\rm yr}})}^{-1/2}\qty(\frac{d_{\rm eff}}{10\,\rm Gpc})^{1/2}\,.
\end{equation}
Currently, the length scale of the matter spectrum accessible through measurements is well above the parsec scale, leaving a vast unexplored range even when accounting for the finite observation time.


\subsection{Observability}\label{Sec: new quantity observability}

In this subsection, we discuss the observability of $I(\omega)$ and its correlation function introduced in Sec. \ref{Sec: new quantity formalization}.
First, a direct observable is the lensed waveform of GWs, including noise: $\tilde{\phi}_{\rm obs}=F\tilde{\phi}_{0}+n$, where $n$ represents the noise.
Hereafter, $\tilde{\phi}_{\rm obs}$ and ${\tilde \phi}_0$ is abbreviated as $\phi_{\rm obs}$ and $\phi_0$, respectively.
To extract the amplification factor from $\phi_{\rm obs}$, we assume a specific wave source and characterize it by the parameters to make an unlensed waveform template $\hat{\phi}_0$, and obtain the observational amplification factor $\hat{F}$ as
\begin{equation}
    \hat{F}=\frac{\phi_{\rm obs}}{\hat{\phi}_0}=F+\delta F_t+\delta F_n\,,\label{Eq: F observational}
\end{equation}
where
\begin{equation}
    \delta F_t=\qty(\frac{\phi_0}{\hat{\phi}_0}-1)F\,,~~\delta F_n=\frac{n}{\hat{\phi}_0}\,.\label{Eq: delta F}
\end{equation}
Substituting Eq. (\ref{Eq: F observational}) into Eq. (\ref{Eq: new quantity def}) yields
\begin{equation}
\label{decomposition-of-I}
    \hat{I}=I+\delta I_t + \delta I_n\,,
\end{equation}
where
\begin{equation}
    \delta I_{t/n}=-\omega^2\frac{d}{d\omega}\frac{\delta F_{t/n}}{\omega}\,.\label{Eq: delta I}
\end{equation}

Let us estimate $\delta I_t$ assuming $\phi_0$ is a quasi-circular orbit binary GW.
Frequency dependence of $\phi_0$ can be expressed using real constants $A$, $B$, $C$ and $D$ as\cite{maggiore2007gravitational}
\begin{equation}
    \phi_0(\omega)=A\omega^{-7/6}e^{i(B\omega^{-5/3}+C\omega+D)}\,.
\end{equation}
Although the true source and the assumed source may, in general, be of different types, in the following analysis, we restrict ourselves to the case where they are of the same type, differing only in their source parameters.
Then, $\hat{\phi}_0$ will take the form as
\begin{equation}
    \hat{\phi}_0(\omega)={\hat A}\omega^{-7/6}e^{i({\hat B}\omega^{-5/3}+{\hat C}\omega+{\hat D})}\,,
\end{equation}
where the parameters ${\hat A}, {\hat B}, {\hat C}$ and ${\hat D}$ will differ from their true values in general.
Since the deviation of $\hat{\phi}_0$ from $\phi_0$ is caused by the lensing effect and the noise, $\phi_0/\hat{\phi}_0-1$ is expected to be $\mathcal{O}(\max(1/\text{SNR},\,|F-1|))\ll\mathcal{O}(1)$. Thus we can approximate as $\phi_0/\hat{\phi}_0-1\simeq a_t+i(b_t\omega^{-5/3}+c_t\omega)$ using new constants $a_t$, $b_t$ and $c_t$ given by
\begin{equation}
    a_t=\frac{A}{\hat A}-1+i(D-{\hat D})\,,~~b_t=B-{\hat B}\,,~~c_t=C-{\hat C}\,.
\end{equation}
Using this approximation and assuming that the lensing effect is weak, i.e., the deviation of $F$ from unity is small, Eq. (\ref{Eq: delta F}) reduces to
\begin{equation}
    \delta F_t\simeq a_t + i(b_t\omega^{-5/3} + c_t\omega)\,,
\end{equation}
and we obtain
\begin{equation}
    \delta I_t \simeq a_t + \frac{8i}{3}b_t\omega^{-5/3}\label{Eq: delta I approx}
\end{equation}
from Eq. (\ref{Eq: delta I}).
While the first term in Eq. (\ref{Eq: delta I approx}) is a constant, the second term is a monotonically decreasing function of frequency, and its maximum value within the observational frequency range $[\omega_{\rm min},\,\omega_{\rm max}]$ is expected to be same order as $a_t$.
Accordingly, we estimate
\begin{equation}
    b_t\omega^{-5/3}=\mathcal{O}\qty(a_t\qty(\frac{\omega_{\rm min}}{\omega})^{5/3})\,,
\end{equation}
where $a_t=\mathcal{O}(\max(1/\text{SNR},\,|F-1|))$.
When the frequency range is sufficiently wide, $\omega$ can be chosen such that the second term in Eq. (\ref{Eq: delta I approx}) becomes much smaller than the first term.
Hence, we neglect the second term and approximate $\delta I_t$ as
\begin{equation}
    \delta I_t\simeq a_t\,.
\end{equation}

When $I(\omega)$ is decomposed into a multiple components like Eq.~(\ref{decomposition-of-I}), there appear several terms in $\langle\hat{I}^*(\omega)\hat{I}(\omega')\rangle$.
First, $\langle I^*(\omega)\delta I_n(\omega')\rangle$ vanishes, due to the statistical independence of the lensing effect and the noise.
Second, $\langle \delta I_n^*(\omega)\delta I_n(\omega')\rangle$ also vanishes under the assumption that the noise is stationary.
Third, $\langle I^*(\omega)a_t\rangle$ does not generally vanish.
For example, if $|F-1|$ is much larger than $1/\text{SNR}$, the deviation of $\hat{\phi}_0$ from $\phi_0$ is mainly due to the lensing effect, and $a_t$ depends on the amplification factor.
However, since $a_t$ is determined by $F(\omega)$ over the entire frequency range $[\omega_{\rm min},\,\omega_{\rm max}]$, its correlation with $F(\omega)$ at a particular $\omega$ is small compared to $\langle a_t^2\rangle$, and therefore can be neglected.
By the same argument, $\langle \delta I_n^*(\omega)a_t\rangle$ can also be neglected.
From the above, we obtain
\begin{equation}
    \langle\hat{I}^*(\omega)\hat{I}(\omega')\rangle\simeq\langle I^*(\omega)I(\omega')\rangle + \langle a_t^2\rangle\,.\label{Eq: correlation obs}
\end{equation}
Here, because the second term is a frequency-independent constant, it does not affect the frequency dependence of $\langle I^*(\omega)I(\omega')\rangle$.
Consequently, the information of the power spectrum can be extracted from $\langle\hat{I}^*(\omega)\hat{I}(\omega')\rangle$.
Note that Eq. (\ref{Eq: correlation obs}) is valid only for a sufficiently large number of samples used for ensemble averaging.
If the sample size is not large enough, an additional noise term may dominate the right-hand side of Eq. (\ref{Eq: correlation obs}).
In this sense, Eq. (\ref{Eq: correlation obs}) implies that even if $I(\omega)$ itself is smaller than the noise, the frequency dependence of its correlation function can in principle be measured by averaging over a large number of samples.

At the end of this section, we also mention an alternative method for measuring the amplification factor, distinct from the discussion above: a lens model-independent method based on dynamic programming\cite{Dai:2018enj}.
In this method, the measurement error is order of $1/\text{SNR}$, so comparing the magnitude of $I(\omega)$ with $1/\text{SNR}$ would be useful.
We then estimate $\sqrt{\langle|I(\omega)|^2\rangle}$ using the one-halo term in the halo model, which dominates the small-scale power spectrum below $1 \text{ Mpc}$—scales of particular interest in this study.
As shown in Appendix \ref{Sec: appendix esimation of I}, $\sqrt{\langle |I^{\rm 1h}(\omega)|^2\rangle}\simeq 10^{-2}$ when $\chi_s=3 \text{ Gpc}$.
On the other hand, the noise level in LISA is roughly $1/\text{SNR}=\mathcal{O}(10^{-3})$\cite{Babak:2021mhe}.
Therefore, the magnitude of $I(\omega)$ is expected to exceed $1/\text{SNR}$ and that $I(\omega)$ will be measurable in future observations.


\section{Conclusion}\label{Sec: conclusion}

We have discussed gravitational lensing of GWs as a means of probing the density distribution of the universe.
In this context, wave optics plays an important role, making the amplification factor sensitive to the structure larger than the Fresnel scale $r_F$.
We then introduced a new quantity, $I(\omega)$, defined in Eq. (\ref{Eq: new quantity def}). Remarkably, the correlation function of $I(\omega)$ between two nearby frequencies defined in Eq. (\ref{Eq: correlation}) can probe an arbitrarily small-scale structure, despite the observational upper bound on the frequency range.
Furthermore, even when accounting for the observational error in the amplification factor, information about the matter power spectrum can be obtained from the frequency dependence of the correlation function.
This will allow us to explore the unknown small-scale structure in the future, revealing the nature of dark matter.


\appendix


\section{Relation Between $I$ and the Convergence Field}
\label{appendix-A}

In \cite{Choi:2021bkx}, it was shown that, for a weakly lensed signal caused by a localized object for which the thin-lens approximation may be used, the derivative of the amplification factor is related to the tangential shear $\gamma_t$ averaged over the circle with its radius given by the Fresnel scale:
\begin{equation}
    \frac{dF(w)}{d\ln w}\simeq\bigg\langle\gamma_t\qty(\frac{1}{\sqrt{w}}e^{\pi i/4})\bigg\rangle_{\phi}\,.\label{Eq: dF/dlnw}
\end{equation}
Here $w$ and $\bm{x}$ are the dimensionless frequency and position within the lens plane, respectively.
In this appendix, we show that a similar relation exists for the newly defined quantity $I(\omega)$.
To this end, we start from
\begin{equation}
    F(w)-1\simeq\overline{\kappa}\qty(\frac{1}{\sqrt{w}}e^{\pi i/4})\,,
\end{equation}
where 
\begin{equation}
    \overline{\kappa}(x)=\frac{1}{\pi x^2}\int_0^{2\pi}d\phi'\int_0^x dx'\,\kappa(\bm{x'})
\end{equation}
is the convergence field averaged within a circle of radius $x$.
From the above equation and $d\overline{\kappa}(x)/d\ln x=-2\langle\gamma_t(x)\rangle_{\phi}$, we have
\begin{equation}
    \overline{\kappa}(x)=\langle{\kappa}(x)\rangle_\phi+\langle\gamma_t(x)\rangle_\phi\,.\label{Eq: kappa decompose}
\end{equation}
Here, from the definition of $I(w)$ in Eq. (\ref{Eq: new quantity def}),
\begin{equation}
    F(w)-1=I(w)+\frac{dF(w)}{d\ln w}\label{Eq: F decompose}
\end{equation}
holds, and comparing Eq. (\ref{Eq: kappa decompose}) and (\ref{Eq: F decompose}) shows that
\begin{equation}
    I(w)\simeq\bigg\langle\kappa\qty(\frac{1}{\sqrt{w}}e^{\pi i/4})\bigg\rangle_{\phi}\,.
\end{equation}
This equation indicates that $I(w)$ is related to the convergence field averaged over the circle with its radius given by the Fresnel scale.


\section{Estimation of $\langle |I|^2\rangle$ in the Halo Model}\label{Sec: appendix esimation of I}

This section roughly estimates the contribution of the one-halo term in the halo model to $\langle|I|^2\rangle$.
The quantity we calculate is 
\begin{eqnarray}
    \langle |I^{\rm 1h}(\omega)|^2\rangle
    &=&\int_0^{\chi_s} d\chi\int \frac{d^2k_{\perp}}{(2\pi)^2}\,\frac{P^{\rm 1h}_{\delta\rho}(k_\perp;\chi)}{\Sigma_c^2(\chi)}\nonumber\\
    &\simeq&\chi_s\int \frac{d^2k_{\perp}}{(2\pi)^2}\,\frac{P^{\rm 1h}_{\delta\rho}(k_\perp;\chi_l)}{\Sigma_c^2(\chi_l)}\int_0^{\chi_s} \frac{d\chi}{\chi_s}\,\frac{\Sigma_c^2(\chi_l)}{\Sigma_c^2(\chi)}\nonumber\\
    &\simeq&\frac{8}{15}\frac{\chi_s}{\Sigma_c^2(\chi_l)}\int \frac{d^2k_{\perp}}{(2\pi)^2}\,P^{\rm 1h}_{\delta\rho}(k_\perp;\chi_l)\,.\label{Eq: I 1h}
\end{eqnarray}
Here, we approximate $P^{\rm 1h}_{\delta\rho}(k_\perp;\chi)$ and $a(\chi)$ by evaluating them at $\chi_l=\chi_s/2$, where $d_{\rm eff}(\chi)$ reaches its maximum, assuming that thair variation with respect to $\chi$ is small.
The one-halo power spectrum is
\begin{equation}
    P^{\rm 1h}_{\delta\rho}(k_\perp;\chi)=\int_0^\infty dM\,M^2n(M, \chi)\,\frac{|\tilde{u}(k_\perp|M,\chi)|^2}{a^3(\chi)}\,,
\end{equation}
where $n(M,\chi)$ is the mass function and $\tilde{u}(k|M,\chi)$ is the halo profile normalized as $\tilde{u}(0|M,\chi)=1$.
Note that the $k_\perp$ is the comoving wave number, while densities such as $\rho$ and $n$ are defined in terms of the physical distance.
For simplicity, the profile is assumed to be Gaussian
\begin{equation}
    \tilde{u}(k|M,\chi)=e^{-k^2R^2_{\rm vir}/2a^2}\,.
\end{equation}
$R_{\rm vir}$ is the virial radius defined by
\begin{equation}
    R_{\rm vir}(M,\chi)=\qty(\frac{3M}{4\pi\Delta_{\rm vir}\rho_m(\chi)})^{1/3}\,,
\end{equation}
where $\Delta_{\rm vir}\simeq178$ and $\rho_m(\chi)$ is the average mass density.
Then we obtain
\begin{equation}
    \int\frac{d^2k_{\perp}}{(2\pi)^2}\,P^{\rm 1h}_{\delta\rho}(k_\perp;\chi)=\frac{1}{4\pi a(\chi)}\int_0^\infty dM\,\frac{M^2n(M, \chi)}{R_{\rm vir}^2(M,\chi)}\,.\label{Eq: int P 1h}
\end{equation}
As the mass function, we use the Press-Schechter mass function:
\begin{equation}
    n(M,\chi)=\sqrt{\frac{2}{\pi}}\frac{\rho_m(\chi)}{M^2}\frac{d\ln\nu}{d\ln M}\nu e^{-\nu^2/2}\,,
\end{equation}
where
\begin{equation}
    \nu(M, \chi)=\frac{\delta_c}{\sigma(M,\chi)}\,.\label{Eq: nu mass func}
\end{equation}
Here, $\delta_c\simeq 1.69$ and $\sigma(M,\chi)$ is the variance of the overdensity.
We approximate $\chi$-dependence of $\sigma(M, \chi)$ by its time evolution in de Sitter space-time: $\sigma(M,\chi)\simeq a(\chi)\sigma(M)$.
Furthermore, by approximating the linear matter power spectrum as $P^{\rm lin}(k)\propto k$, mass dependence of $\sigma(M)$ becomes $\sigma(M)\propto M^{-2/3}$.
This proportional coefficient is determined by the relation
\begin{equation}
    \sigma(M)=\sigma_8\qty(\frac{R_8}{R(M)})^2\,,~~R(M)=\qty(\frac{3M}{4\pi\rho_{m0}})^{1/3}\,,\label{Eq: R mass func}
\end{equation}
where $\sigma_8\simeq0.829$, $R_8\simeq 11.4\text{ Mpc}$, and $\rho_{m0}\simeq9.52\times10^{-8}M_{\odot}/\text{pc}^3$.
From Eq. (\ref{Eq: nu mass func}) and (\ref{Eq: R mass func}), we can write
\begin{equation}
    \nu(M,\chi)=\qty(\frac{M}{M_c})^{2/3}=\qty(\frac{R_{\rm vir}(M,\chi)}{R_c(\chi)})^2\,,
\end{equation}
where
\begin{equation}
    M_c(\chi)=a^{3/2}(\chi)M_*\,,~~M_*=\qty(\frac{\sigma_8}{\delta_c})^{3/2}\frac{4}{3}\pi R_8^3\rho_{m0}\simeq2.05\times10^{14}M_{\odot}
\end{equation}
and
\begin{equation}
    R_c(\chi)=a^{3/2}(\chi)R_*\,,~~R_*=\Delta_{\rm vir}^{-1/3}\qty(\frac{\sigma_8}{\delta_c})^{1/2}R_8\simeq 1.43\text{ Mpc}\,.
\end{equation}
From above, Eq. (\ref{Eq: int P 1h}) becomes
\begin{equation}
    \int\frac{d^2k_{\perp}}{(2\pi)^2}\,P^{\rm 1h}_{\delta\rho}(k_\perp;\chi)=\frac{A}{4\pi}a^{-9/2}(\chi)\frac{\rho_{m0}M_*}{R_*^2}\,,\label{Eq: int P 1h 2nd}
\end{equation}
where 
\begin{equation}
    A=\sqrt{\frac{2}{\pi}}\int_0^\infty d\nu\,\sqrt{\nu}e^{-\nu^2/2}\simeq 0.822\,.
\end{equation}
Finally, substituting Eq. (\ref{Eq: int P 1h 2nd}) and
\begin{equation}
    \Sigma_c(\chi_l)=\frac{1}{\pi G \chi_s a^2(\chi_l)}\simeq \frac{2.22\times10^3}{a^2(\chi_l)}\frac{M_\odot}{\text{pc}^2}\frac{3\text{ Gpc}}{\chi_s}
\end{equation}
and $a(\chi_l=1.5\text{ Gpc})\simeq0.698$ into Eq. (\ref{Eq: I 1h}) gives
\begin{equation}
    \sqrt{\langle |I^{\rm 1h}(\omega)|^2\rangle}\simeq 1.57\times 10^{-2}\qty(\frac{\chi_s}{3\text{ Gpc}})^{3/2}\,.
\end{equation}
%


\section*{Acknowledgements}
This work was supported by JSPS KAKENHI Grant Number 24KJ1094 (ST) and by JSPS KAKENHI Grant Nos. JP23K03411 (TS).
\bibliography{ref}

@article{Tambalo:2022wlm,
    author = "Tambalo, Giovanni and Zumalac\'arregui, Miguel and Dai, Liang and Cheung, Mark Ho-Yeuk",
    title = "{Gravitational wave lensing as a probe of halo properties and dark matter}",
    eprint = "2212.11960",
    archivePrefix = "arXiv",
    primaryClass = "astro-ph.CO",
    doi = "10.1103/PhysRevD.108.103529",
    journal = "Phys. Rev. D",
    volume = "108",
    number = "10",
    pages = "103529",
    year = "2023"
}

@article{Savastano:2023spl,
    author = "Savastano, Stefano and Tambalo, Giovanni and Villarrubia-Rojo, Hector and Zumalacarregui, Miguel",
    title = "{Weakly lensed gravitational waves: Probing cosmic structures with wave-optics features}",
    eprint = "2306.05282",
    archivePrefix = "arXiv",
    primaryClass = "gr-qc",
    doi = "10.1103/PhysRevD.108.103532",
    journal = "Phys. Rev. D",
    volume = "108",
    number = "10",
    pages = "103532",
    year = "2023"
}

@article{Oguri:2020ldf,
    author = "Oguri, Masamune and Takahashi, Ryuichi",
    title = "{Probing Dark Low-mass Halos and Primordial Black Holes with Frequency-dependent Gravitational Lensing Dispersions of Gravitational Waves}",
    eprint = "2007.01936",
    archivePrefix = "arXiv",
    primaryClass = "astro-ph.CO",
    doi = "10.3847/1538-4357/abafab",
    journal = "Astrophys. J.",
    volume = "901",
    number = "1",
    pages = "58",
    year = "2020"
}

@article{Takahashi:2005ug,
    author = "Takahashi, Ryuichi",
    title = "{Amplitude and phase fluctuations for gravitational waves propagating through inhomogeneous mass distribution in the universe}",
    eprint = "astro-ph/0511517",
    archivePrefix = "arXiv",
    doi = "10.1086/503323",
    journal = "Astrophys. J.",
    volume = "644",
    pages = "80--85",
    year = "2006"
}

@article{Moore:1999nt,
    author = "Moore, B. and Ghigna, S. and Governato, F. and Lake, G. and Quinn, Thomas R. and Stadel, J. and Tozzi, P.",
    title = "{Dark matter substructure within galactic halos}",
    eprint = "astro-ph/9907411",
    archivePrefix = "arXiv",
    doi = "10.1086/312287",
    journal = "Astrophys. J. Lett.",
    volume = "524",
    pages = "L19--L22",
    year = "1999"
}

@article{Lin:2019uvt,
    author = "Lin, Tongyan",
    title = "{Dark matter models and direct detection}",
    eprint = "1904.07915",
    archivePrefix = "arXiv",
    primaryClass = "hep-ph",
    doi = "10.22323/1.333.0009",
    journal = "PoS",
    volume = "333",
    pages = "009",
    year = "2019"
}

@article{Takahashi:2005sxa,
    author = "Takahashi, Ryuichi and Suyama, Teruaki and Michikoshi, Shugo",
    title = "{Scattering of gravitational waves by the weak gravitational fields of lens objects}",
    eprint = "astro-ph/0503343",
    archivePrefix = "arXiv",
    doi = "10.1051/0004-6361:200500140",
    journal = "Astron. Astrophys.",
    volume = "438",
    pages = "L5",
    year = "2005"
}

@article{Tambalo:2022plm,
    author = "Tambalo, Giovanni and Zumalac\'arregui, Miguel and Dai, Liang and Cheung, Mark Ho-Yeuk",
    title = "{Lensing of gravitational waves: Efficient wave-optics methods and validation with symmetric lenses}",
    eprint = "2210.05658",
    archivePrefix = "arXiv",
    primaryClass = "gr-qc",
    doi = "10.1103/PhysRevD.108.043527",
    journal = "Phys. Rev. D",
    volume = "108",
    number = "4",
    pages = "043527",
    year = "2023"
}

@article{Dai:2018enj,
    author = "Dai, Liang and Li, Shun-Sheng and Zackay, Barak and Mao, Shude and Lu, Youjun",
    title = "{Detecting Lensing-Induced Diffraction in Astrophysical Gravitational Waves}",
    eprint = "1810.00003",
    archivePrefix = "arXiv",
    primaryClass = "gr-qc",
    doi = "10.1103/PhysRevD.98.104029",
    journal = "Phys. Rev. D",
    volume = "98",
    number = "10",
    pages = "104029",
    year = "2018"
}

@article{Tanaka:2023mvy,
    author = "Tanaka, So and Suyama, Teruaki",
    title = "{Kramers-Kronig relation in gravitational lensing}",
    eprint = "2303.05650",
    archivePrefix = "arXiv",
    primaryClass = "gr-qc",
    doi = "10.1103/PhysRevD.108.044015",
    journal = "Phys. Rev. D",
    volume = "108",
    number = "4",
    pages = "044015",
    year = "2023"
}

@article{Nakamura:1999uwi,
    author = "Nakamura, Takahiro T. and Deguchi, Shuji",
    title = "{Wave Optics in Gravitational Lensing}",
    doi = "10.1143/ptps.133.137",
    journal = "Prog. Theor. Phys. Suppl.",
    volume = "133",
    pages = "137--153",
    year = "1999"
}

@article{Mizuno:2022xxp,
    author = "Mizuno, Morifumi and Suyama, Teruaki",
    title = "{Weak lensing of gravitational waves in wave optics: Beyond the Born approximation}",
    eprint = "2210.02062",
    archivePrefix = "arXiv",
    primaryClass = "astro-ph.CO",
    doi = "10.1103/PhysRevD.108.043511",
    journal = "Phys. Rev. D",
    volume = "108",
    number = "4",
    pages = "043511",
    year = "2023"
}

@article{peters1974index,
  title={Index of refraction for scalar, electromagnetic, and gravitational waves in weak gravitational fields},
  author={Peters, Philip C},
  journal={Physical Review D},
  volume={9},
  number={8},
  pages={2207},
  year={1974},
  publisher={APS}
}

@book{Peebles:1980yev,
    author = "Peebles, P. James",
    title = "{The Large-Scale Structure of the Universe}",
    isbn = "978-0-691-08240-0, 978-0-691-20983-8, 978-0-691-20671-4",
    publisher = "Princeton University Press",
    month = "11",
    year = "1980"
}

@article{Kim:2025njb,
    author = "Kim, Sungjung and Gil Choi, Han and Jung, Sunghoon",
    title = "{Probing small-scale power spectrum with gravitational-wave diffractive lensing}",
    eprint = "2501.14904",
    archivePrefix = "arXiv",
    primaryClass = "hep-ph",
    month = "1",
    year = "2025"
}

@article{Choi:2021bkx,
    author = "Choi, Han Gil and Park, Chanung and Jung, Sunghoon",
    title = "{Small-scale shear: Peeling off diffuse subhalos with gravitational waves}",
    eprint = "2103.08618",
    archivePrefix = "arXiv",
    primaryClass = "astro-ph.CO",
    doi = "10.1103/PhysRevD.104.063001",
    journal = "Phys. Rev. D",
    volume = "104",
    number = "6",
    pages = "063001",
    year = "2021"
}

@article{Nakamura:1997sw,
    author = "Nakamura, Takahiro T.",
    title = "{Gravitational lensing of gravitational waves from inspiraling binaries by a point mass lens}",
    reportNumber = "UTAP-272-97, YITP-97-61",
    doi = "10.1103/PhysRevLett.80.1138",
    journal = "Phys. Rev. Lett.",
    volume = "80",
    pages = "1138--1141",
    year = "1998"
}

@article{Takahashi:2003ix,
    author = "Takahashi, Ryuichi and Nakamura, Takashi",
    title = "{Wave effects in gravitational lensing of gravitational waves from chirping binaries}",
    eprint = "astro-ph/0305055",
    archivePrefix = "arXiv",
    doi = "10.1086/377430",
    journal = "Astrophys. J.",
    volume = "595",
    pages = "1039--1051",
    year = "2003"
}

@article{Babak:2021mhe,
    author = "Babak, Stanislav and Petiteau, Antoine and Hewitson, Martin",
    title = "{LISA Sensitivity and SNR Calculations}",
    eprint = "2108.01167",
    archivePrefix = "arXiv",
    primaryClass = "astro-ph.IM",
    reportNumber = "LISA-LCST-SGS-TN-001",
    month = "8",
    year = "2021"
}

@book{maggiore2007gravitational,
  title={Gravitational waves: Volume 1: Theory and experiments},
  author={Maggiore, Michele},
  year={2007},
  publisher={OUP Oxford}
}

\end{document}